\documentclass[12pt]{iopart}
\usepackage{graphicx}
\begin{document}

\title[Solitons in Tonks-Girardeau gas with dipolar interactions]
{Solitons in Tonks-Girardeau gas with dipolar interactions}
\author{B B Baizakov$^{1}$, F Kh Abdullaev$^{1}$, B A Malomed$^{2}$ and \\ M Salerno$^{3}$}
\address{$^1$ Physical - Technical Institute, Uzbek Academy of Sciences, 2-b, G.
Mavlyanov str., 100084, Tashkent, Uzbekistan \\
$^2$ Department of Physical Electronics, School of Electrical Engineering,
Faculty of Engineering, Tel Aviv University, Tel Aviv 69978, Israel \\
$^3$ Dipartimento di Fisica ``E. R. Caianiello'', Consorzio
Nazionale Interuniversitario per le Scienze Fisiche della Materia
(CNISM), Universit\'a di Salerno, I-84081, Baronissi (SA), Italy }
\ead{baizakov@uzsci.net, fatkh@uzsci.net, malomed@eng.tau.ac.il,
salerno@sa.infn.it}

\date{\today}

\begin{abstract}
The existence of bright solitons in the model of the Tonks-Girardeau
(TG) gas with dipole-dipole (DD) interactions is reported. The
governing equation is taken as the quintic nonlinear Schr\"{o}dinger
equation (NLSE)\ with the nonlocal cubic term accounting for the DD
attraction. In different regions of the parameter space (the dipole
moment and atom number), matter-wave solitons feature flat-top or
compacton-like shapes. For the flat-top states, the NLSE with the
local cubic-quintic (CQ) nonlinearity is shown to be a good
approximation. Specific dynamical effects are studied assuming that
the strength of the DD interactions is ramped up or drops to zero.
Generation of dark-soliton pairs in the gas shrinking under the
action of the intensifying DD attraction is observed. Dark solitons
exhibit the particle-like collision behavior. Peculiarities of
dipole solitons in the TG gas are highlighted by comparison with the
NLSE including the local CQ terms. Collisions between the solitons
are studied too. In many cases, the collisions result in merger of
the solitons into a breather, due to strong attraction between them.
\end{abstract}

 \submitto{\JPB} \maketitle

\section{Introduction}

The experimental realization of degenerate Bose gases in tight
effectively one-dimensional (1D) traps \cite{paredes} in the
Tonks-Girardeau (TG) regime \cite{tg} has rekindled interest to
the TG model, which, being well known for a long time, was assumed
to have a theoretical value only (for a review see
\cite{yukalov}). In the TG state, bosons emulate the Pauli
exclusion principle, as a result of the hard-core repulsion,
rather than as a manifestation of the quantum statistics. One of
recent trends in this field is the application of mean-field-like
approaches to the description of macroscopic dynamics of TG gases
\cite{kolomeisky,damski}. The starting point of this approach is
the use of a formal analogy between hydrodynamic equations for
degenerate Fermi gases and Bose gases in the TG phase. This
analogy has led to the derivation of the nonlinear Schr\"{o}dinger
equation (NLSE) with the local quintic repulsive nonlinearity for
the ``wave function" of the TG gas \cite{kolomeisky}. In the
framework of the quintic NLSE, the existence of dark solitons was
predicted \cite{kolomeisky,bhaduri}, and dynamics of dark solitons
was investigated \cite{frantzeskakis}. Bright solitons of the gap
type in the same model equation including a periodic
optical-lattice (OL) potential have been reported too, although in
different contexts, such as a phenomenological description of
degenerate Fermi gases and BCS superfluids,
\cite{A-S05}-\cite{Lisbon}. The quasi-mean-field approach relies
on the physically plausible assumption that the spatial scale of
variations of the gas density is much larger than the healing
length. Under this condition, the effective NLSE allows one to
approximate collective oscillations of the TG gas in a harmonic
trap. Indeed, it has been demonstrated that excitation frequencies
derived from the fermionic hydrodynamic equations, and from the
quintic NLSE agree within a few percent \cite{Minguzzi,AG}.

A natural extension of this line of research is to consider gases with
long-range dipole-dipolar (DD) interactions between atoms. Some bosonic
atoms, such as $^{52}$Cr, feature a significant permanent magnetic dipole
moment ($\simeq 6\mu _{B}$, in the case of chromium). The creation of the
Bose-Einstein condensate (BEC) made of $^{52}$Cr, and various experiments in
that quantum gas have been reported \cite{Pfau,Ueda,griesmaier}. A gas of
LiCs molecules carrying a permanent electric dipole moment was also recently
made available to experiments \cite{LiCs}. In addition to that, atoms may be
polarized by an external dc electric field \cite{marinescu}.

The nonlocal character of the DD interactions may drastically
modify properties of the quantum gas -- first of all, changing the
character of the collapse in it \cite{Ueda}. Further, stable
isotropic \cite{lashkin} and anisotropic \cite{Tikho} 2D solitons
have been predicted in the 2D dipolar BEC, whereas such localized
states are always unstable in the same model with local
interactions \cite{Review}. Recently, families of 1D matter-wave
solitons in a Bose-Einstein condensate supported by the
competition of contact and dipole-dipole interactions of opposite
signs were predicted in Ref. \cite{cuevas}, and 1D discrete
solitons corresponding to the limit case of the dipolar condensate
trapped in a very deep OL (i.e., solutions to the discrete NLSE
with the long-range DD interactions between lattice sites) were
found too \cite{Serbia}. It is relevant to mention that
qualitatively similar effects may be induced by the nonlocal
nonlinearity in models of optical media \cite{bang, kartashov}.

In this work, we study localized structures in the model of a 1D Bose gases
supported by competing nonlocal attractive cubic (DD) and local repulsive
quintic (contact) nonlinearities, using numerical simulations of the NLSE
containing this combination of the nonlinear terms. Besides the cold bosons
in the TG regime, this model may apply (at least, at the phenomenological
level \cite{adhikari-malomed}) to nearly-1D degenerate Fermi gases with DD
interactions between atoms.

The paper is structured as follows. In Section II we introduce the model,
present typical shapes of solitons expected in the TG dipolar gas, and study
their stability in direct simulations. In Section III we demonstrate, via
numerical experiments, dynamics predicted by the model with a \emph{variable}
(time-dependent) strength of the DD interactions, including formation of
dark solitons on top of a broad bright soliton, splitting of the soliton,
and expansion of the TG gas, in cases when the DD interaction is ramped up,
or switched off. In Section IV, we report various results of interactions
and collisions between two solitons. Section V concludes the paper.

\section{The model and numerical analysis}

In accordance with what was said above, we start with the quintic NLSE
introduced in Ref. \cite{kolomeisky}, to which we add the nonlocal cubic
term accounting for the DD interaction:
\begin{equation}
i\psi _{t}+\frac{1}{2}\psi _{xx}-\pi ^{2}N^{2}|\psi |^{4}\psi
+2Nd^{2}\psi (x,t)\int_{-\infty }^{+\infty }R(|x-x^{\prime }|)\
|\psi (x^{\prime },t)|^{2}dx^{\prime }=0, \quad \label{quint}
\end{equation}
where $N$ is the total number of atoms, and $d$ is the atomic
dipole moment, the wave function being subject to the
normalization condition,
\begin{equation}
\int_{-\infty }^{+\infty }\left\vert \psi \left( x\right) \right\vert
^{2}dx=1.  \label{norm}
\end{equation}
Equation (\ref{quint}) is expected to be valid for large number of
atoms $N$, typically $\gg 10$, when oscillations of the
matter-wave density along the chain of strongly interacting bosons
confined to a tight waveguide are essentially suppressed. On the
other hand, the Bose-Fermi mapping, which underlies the solution
of the TG model \cite{tg}, may hold in the present situation
provided that the dipolar attraction between atoms is weaker than
the contact (hard-core) repulsion. These conditions may be
together expressed as $N \gg 2d^{2}/(\pi ^{2}\xi )$, where $\xi $
is the healing length, which is of order one in the present
notation. We assume that the dipoles are oriented along the $x$
axes, hence and the configuration possesses the cylindrical
symmetry.

The model based on Eq. (\ref{quint}) has been studied in detail in
some limit cases. With $R(x)=\delta (x)$, the equation reduces to
the ordinary cubic-quintic (CQ) NLSE, which supports a fully
stable family of bright-soliton solutions in a finite range of the
chemical potential \cite{pushkarov}. On the other hand, if the
quintic term is absent, Eq. (\ref{quint}) reduces to the NLSE with
nonlocal cubic nonlinearity, whose solutions were studied too,
see, e.g., Refs. \cite{mitchell}. The degree of non-locality is
quantified by the ratio between the soliton's width and spatial
extent of nonlocal response function $R(x)$. Depending on this
ratio, qualitatively different behaviors can be observed. Namely,
in the case of strong nonlocality, i.e., when the characteristic
nonlocal response length is greater than the soliton's width, a
sinusoidally ``breathing" mode (the so-called ``accessible
soliton") was predicted \cite{snyder} and experimentally observed
\cite{conti}. In the opposite situation, when the width of the
response function is small compared to the soliton's size, the
governing equation may be approximated by a modified NLSE
\cite{krolikowski2}.

As the kernel in Eq. (\ref{quint}) we use
\begin{equation}
R(x)=\sqrt{\pi }(1+2x^{2})\exp (x^{2})\mathrm{erfc}(|x|)-2|x|,
\label{kernel}
\end{equation}
which was derived for the dipolar BEC in the quasi-1D trap \cite{sinha},
assuming that the dipole moments are fixed (by an external magnetic field)
along axis $x$, hence the DD interaction is attractive.

\begin{figure}[th]
\centerline{\includegraphics[width=5cm, height=5cm]{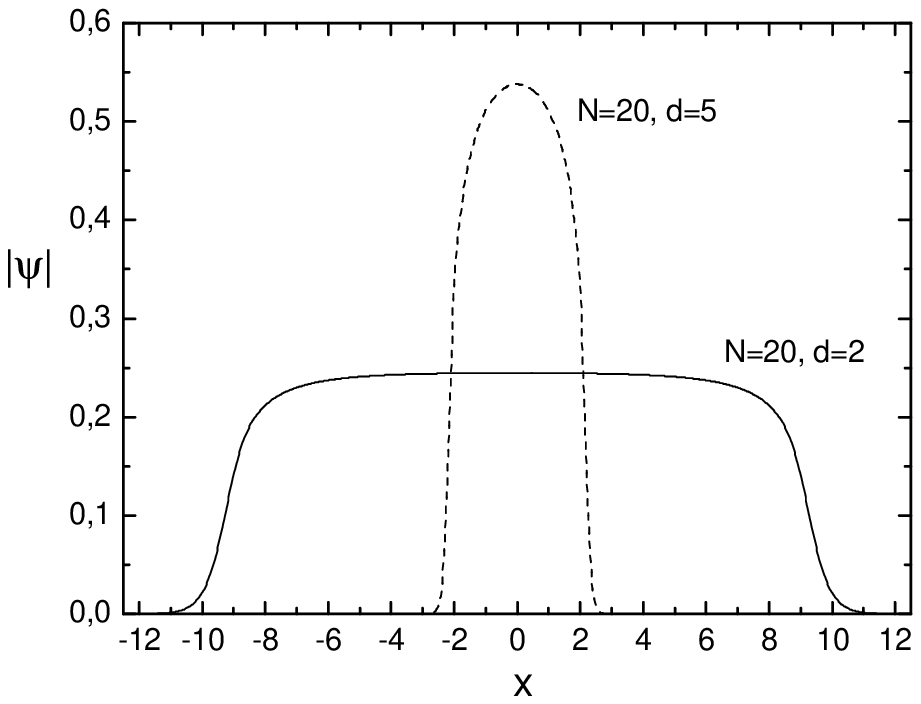}
            \includegraphics[width=5cm, height=5cm]{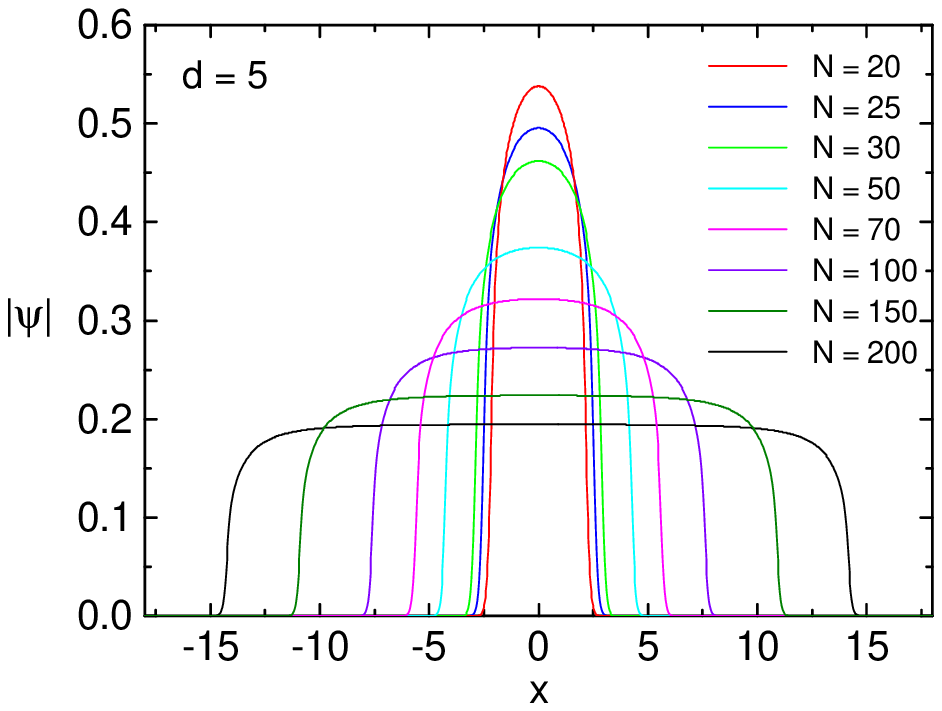}
            \includegraphics[width=5cm, height=5cm]{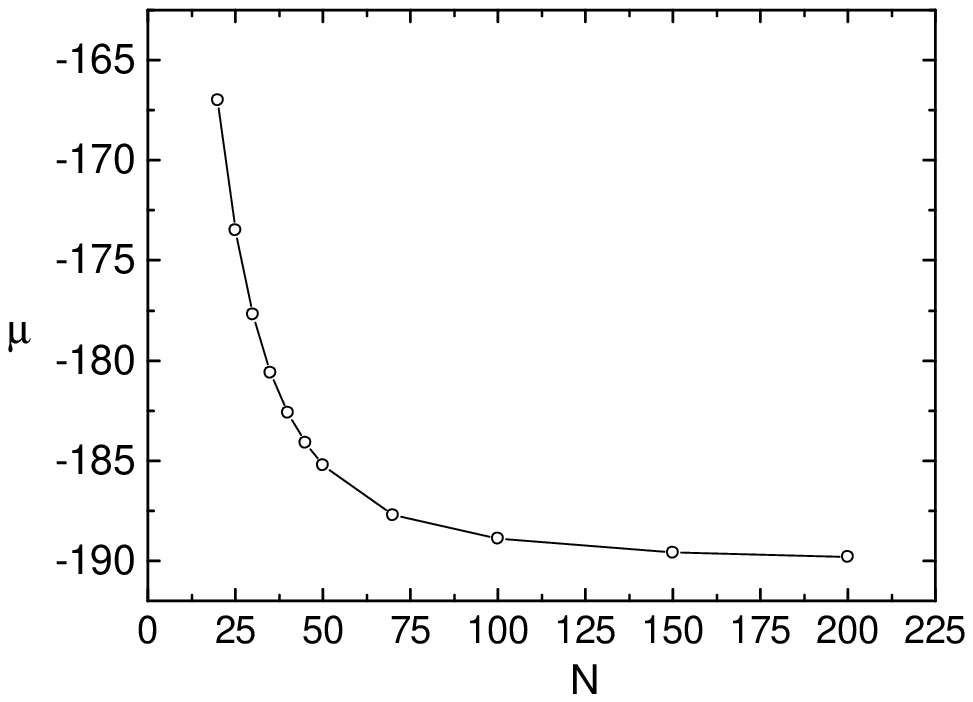}}
\caption{(Color online) Stationary localized states of Eq.
(\protect\ref{quint}), found by means of the numerical integration
in imaginary time. Left panel: Effect of increasing strength of
dipolar interactions at constant number of atoms ($N = 20$).
Soliton is broad (narrow) at weaker (stronger) dipolar
interactions. Middle panel: Effect of increasing number of atoms
at constant strength of dipolar interactions ($d = 5$). Soliton
broadens as the number of atoms increases. Right panel: Chemical
potential as a function the number of atoms for configurations
shown in the middle panel.} \label{fig1}
\end{figure}

As said above, Eq. (\ref{quint}) with the long-range attraction gives rise
to stable solitons in the absence of the quintic repulsive term, a
distinctive feature of these solitons being the breathing intrinsic mode
that may be easily excited \cite{snyder}. Effects produced by the quintic
term in numerically found solutions increase with $N$, leading to broadening
of the localized state, as shown in Fig. \ref{fig1}, which displays soliton
solutions found by means of the imaginary-time relaxation method.

\begin{figure}[th]
\centerline{\includegraphics[width=8cm, height=4cm]{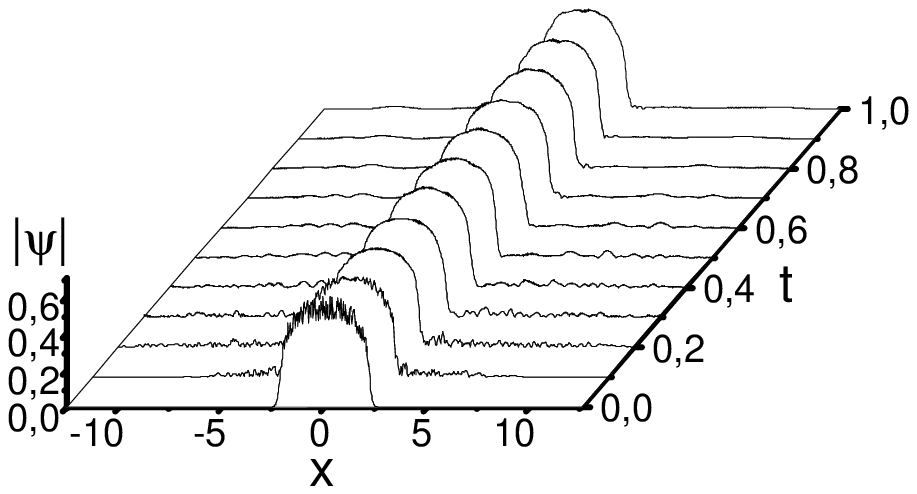}
            \includegraphics[width=8cm, height=4cm]{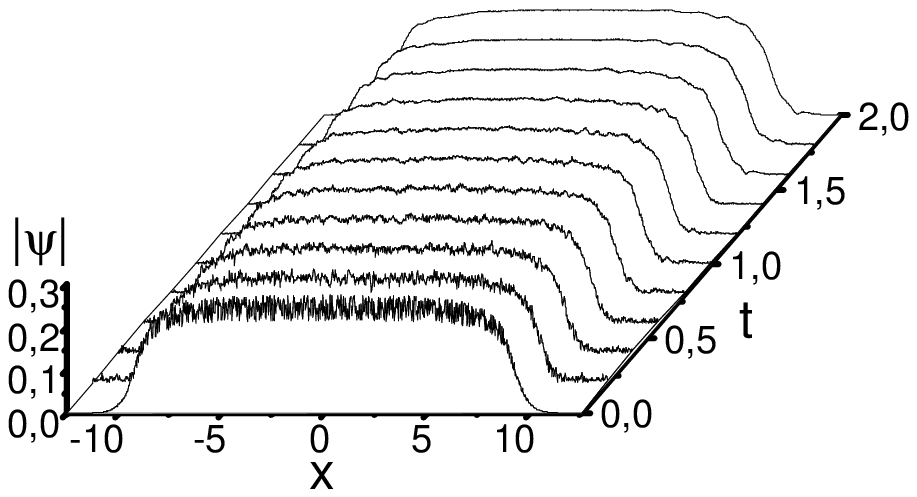}}
\caption{Left panel: The evolution of the soliton with $N=20$ and
$d=5$, depicted in Fig. \protect\ref{fig1} by the dashed line,
which was initially perturbed by adding random noise,
$\protect\psi (x,0)=\protect\psi
_{\mathrm{sol}}(x)[1+\protect\sigma (x)]$, where $\protect\sigma
(x)$ is a random function of $x$, uniformly distributed in
interval $[-0.25,+0.25]$ (note that this perturbation is not
really small). Right panel: A similar verification of the
stability of the flat-top dipole soliton with $N=20$ and $d=2$,
shown by the solid line in Fig. \protect\ref{fig1}.} \label{fig2}
\end{figure}

To check the stability of the localized stationary solutions to
Eq. (\ref{quint}), we ran simulations in the real time, adding
spatially random perturbations to the initial state. The solitons
were observed to shed off the perturbations, in the form of linear
waves, and quickly restored their stationary form, as demonstrated
in Fig. \ref{fig2}, which clearly demonstrates that the localized
states are stable. A first evidence of the stability of solitons
can be seen on the right panel of Fig. \ref{fig1}, where the
chemical potential is drawn as a function of the number of atoms.
Negative slope of the curve ($d\mu/dN < 0$) is the indication of
stability of solitons according to the Vakitov - Kolokolov
criterion \cite{vk}. At large number of atoms ($N > 100$ for
$d=5$), when the soliton acquires a "flat-top" shape, the slope
becomes vanishing and the VK criterion does not apply in a strong
sense (marginal VK stability). For "flat-top" solitons computation
of the chemical potential using the exact solution (see below Eq.
(\ref{exact1})), rather than solution found from the imaginary
time propagation method, indeed gives the constant value $\mu = -
190$ for the whole interval of $N$ shown on the right panel of
Fig. \ref{fig1}.

Despite the indefinite character of the VK criterion in this case,
the direct analysis demonstrates that, in the model with the
competing nonlinearities considered in this work, which correspond
to the attractive nonlocal cubic and repulsive local quintic
terms, all bright solitons turn out to be {\rm stable}. To check
the stability of the localized stationary solutions to Eq.
(\ref{quint}), we ran simulations in the real time, adding
spatially random perturbations to the initial state. The solitons
were observed to shed off the perturbations, in the form of linear
waves, and quickly restored their stationary form, as demonstrated
in Fig. \ref{fig2}, which clearly demonstrates that the localized
states are stable.

All numerical simulations were performed by dint of the split-step
fast-Fourier-transform method \cite{press} in a spatial domain of
length $L=8\pi $ with 1024 modes. The time step was $\delta
t=0.001$. To control the accuracy of numerical results, we
monitored the accuracy of normalization condition (\ref{norm}).
During the entire simulation, it was held to the relative
precision better than $10^{-3}$. To prevent re-entering of the
linear waves emitted by the perturbed soliton into the integration
domain, absorbers were installed at domain boundaries. Actually,
this numerical procedure is a straightforward extension of that
used for simulations of the NLSE in real time \cite{agrawal}, and
imaginary-time propagation method for finding ground states in
NLSE-based models \cite{chiofalo}. In the numerical results
presented below we assumed that the ground state of the
configuration is attained if the variation of the chemical
potential
$$ \mu = \int_{-\infty}^{\infty} \left(\frac{1}{2}|\psi_x|^2 + \pi^2
N^2 |\psi|^6 - 2 d^2 N |\psi|^2 \int_{-\infty }^{+\infty
}R(|x-x^{\prime }|)\ |\psi (x^{\prime },t)|^{2}dx^{\prime }
\right)dx,$$ in the course of integration becomes less than $d\mu
\sim 10^{-8}$.

Shapes of localized states in Fig. \ref{fig1} are determined by
the long-range DD forces and short-range contact repulsion for a
particular number of atoms in the TG gas. Having analyzed a large
body of numerical results, we can conclude that, for a moderate
size of the dipole moment, $d$, the soliton develops a a
``flat-top" shape. In the case of the strong DD interaction, the
soliton becomes a ``compacton", with very short tails. In fact,
the latter case corresponds to the Thomas-Fermi limit, when the
kinetic-energy term in Eq. (\ref{quint}) is negligible, while the
integral term in Eq. (\ref{quint}) plays the role of an effective
potential in which the locally repulsive TG gas is trapped. Then,
stronger DD interactions mean a tighter trapping potential with
steep walls induced by kernel function (\ref{kernel}).

It is pertinent to mention that, in contrast to the recently considered
model of dipolar BEC with competing cubic local and nonlocal interactions
\cite{cuevas}, in the present model, which is of the CQ type, parameter $N$
cannot be eliminated from governing equation (\ref{quint}) by a rescaling of
the wave function. This is the key point explaining the existence of
arbitrarily broad flat-top solitons in the model. On the other hand, if a
broad flat-top bright soliton is available, one may consider it as a
background for dark solitons. The advantage of this setting is a possibility
to explore dark solitons and their interactions on top of the uniform
background, which is distinct from the usual situation when the quantum gas
is confined by a harmonic trap, hence the background matter-wave density is
nonuniform \cite{dark-soliton} (see below).

If the soliton's width greatly exceeds that of kernel
(\ref{kernel}), which takes place at $N>>d^{2}$, Eq. (\ref{quint})
is reduced to the local NLSE with the CQ nonlinearity. Considering
the local counterpart of Eq. (\ref{quint}), we replace
$R(x)\rightarrow 2\delta (x)$, because $\int_{-\infty }^{\infty
}R(x)dx=2$. Then the corresponding local NLSE acquires the form of
\begin{equation}
i\psi _{t}+\frac{1}{2}\psi _{xx}-\alpha |\psi |^{4}\psi +\beta |\psi
|^{2}\psi =0,  \label{locgpe}
\end{equation}
where notation $\alpha =\pi ^{2}N^{2}$, $\beta =4d^{2}N$ is introduced.

\begin{figure}[tbp]
\centerline{\includegraphics[width=8cm, height=6cm]{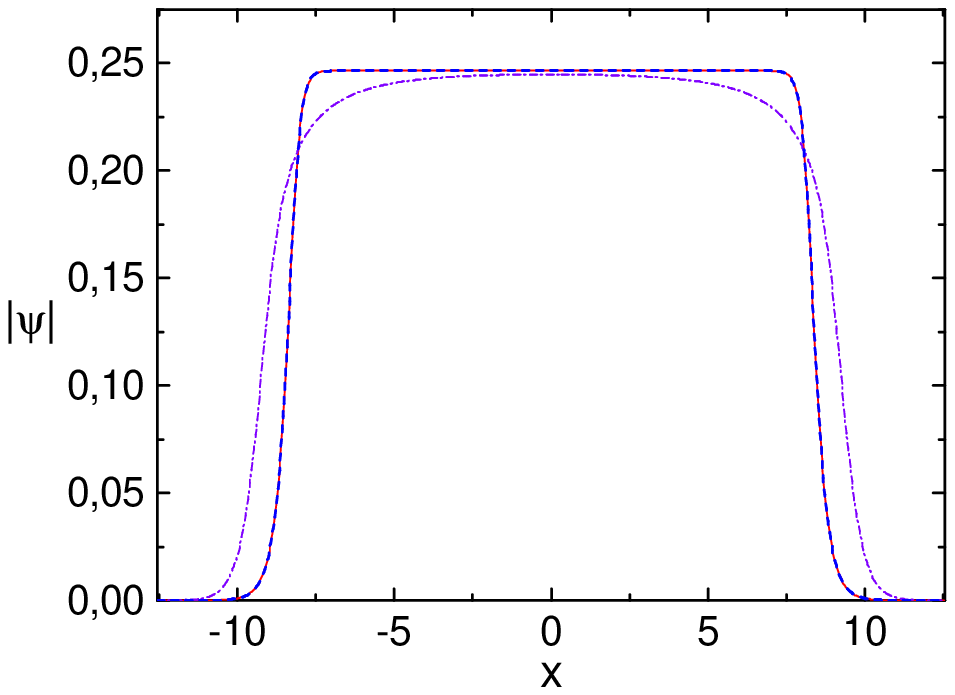}}
\caption{(Color online) Comparison of soliton profiles as
generated by the imaginary-time relaxation method, applied to Eq.
(\protect\ref{quint}) (the dash-dotted purple line) and to the
local equation with the cubic-quintic nonlinearity, Eq.
(\protect\ref{locgpe}) (the solid red line), and as predicted by
exact solution (\protect\ref{exact1}) to the latter equation (the
dashed blue line), for $N=20$, $d=2$. The two latter profiles are
indistinguishable, being juxtaposed to illustrate the accuracy of
the numerical procedure. } \label{fig3}
\end{figure}

Exact soliton solutions to Eq. (\ref{locgpe}) were found in Ref.
\cite{pushkarov}. For the case of the self-focusing cubic ($\beta
> 0$) and defocusing quintic ($\alpha >0$) nonlinearities, the
solution is (under normalization condition (\ref{norm}))
\begin{equation}
\psi (x,t)=\sqrt{\frac{3\beta}{4\alpha}} \ \frac{{\rm tanh}(\eta)
\ \exp[i(qx-\mu t)]}{\sqrt{1 + \mathrm{sech}(\eta)
\mathrm{cosh}(x/a)}}, \quad \eta \equiv \sqrt{\frac{2\alpha}{3}},
\quad a \equiv \frac{1}{\beta}\frac{\eta}{\mathrm{tanh}(\eta)},
\label{exact1}
\end{equation}
where $q$ and $\mu $ stand for the wave vector and chemical
potential of the soliton. An example of stationary solution of Eq.
(\ref{locgpe}) for a particular set of parameters, compared with
the solution of original Eq. (\ref{quint}), is presented in Fig.
\ref{fig3}.

As seen from this figure, the solutions of nonlocal equation
(\ref{quint}) with kernel (\ref{kernel}), and of the local NLSE
are in a qualitative agreement. Therefore, exact solutions
(\ref{exact1}) of the local equation may be used as appropriate
initial conditions for simulations of Eq. (\ref{quint}).

\section{Soliton dynamics under varying dipolar interaction}

Properties of solitons can be studied under a varying strength of the DD
interaction. In the experiments, it may be varied in time by changing the
orientation of dipoles with respect to the axial direction, rotating the
external magnetic field \cite{griesmaier}, or by changing the strength of
the external electric field, if the atomic dipole moment is induced by the
latter field \cite{marinescu}.

Below we present numerical simulations of Eq. (\ref{quint}) with variable
coefficient $d(t)$. With this objective in mind, we first prepare the
ground-state solution of Eq. (\ref{quint}) by means of the imaginary-time
propagation method, as explained in the previous section. Then we insert
this solution into Eq. (\ref{quint}) as the initial condition, and simulate
the evolution in real time, with $d$ substituted by $d(t)$ of a chosen form.

\subsection{Contraction of the TG gas by strengthening the dipolar
interaction}

\begin{figure}[tbp]
\centerline{\includegraphics[width=5cm,height=6cm,clip]{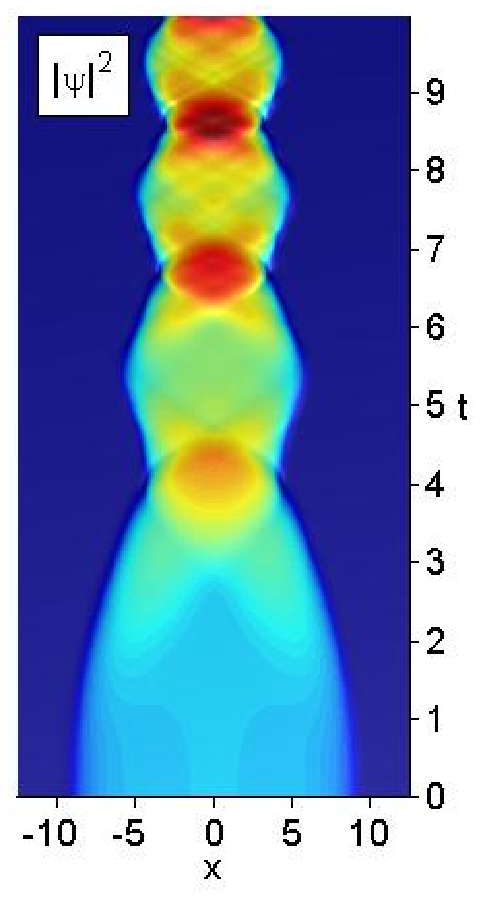}\quad
            \includegraphics[width=5cm,height=6cm,clip]{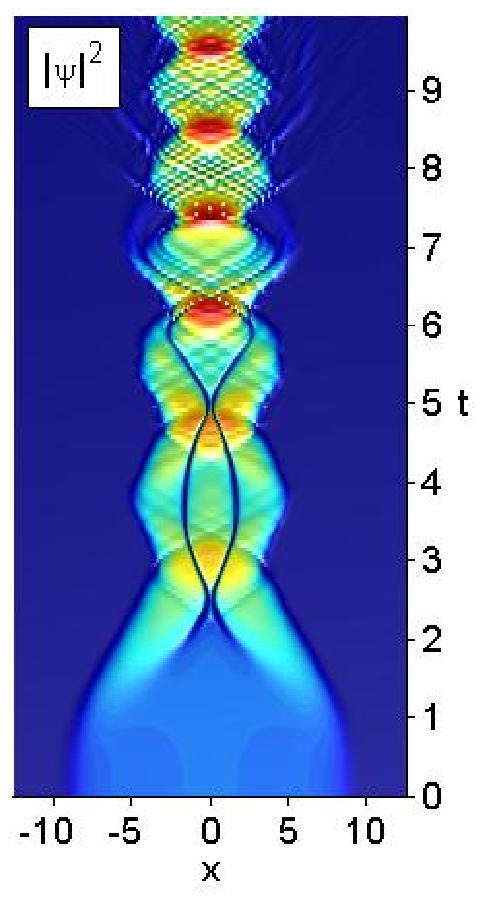}\quad
            \includegraphics[width=5cm,height=6cm,clip]{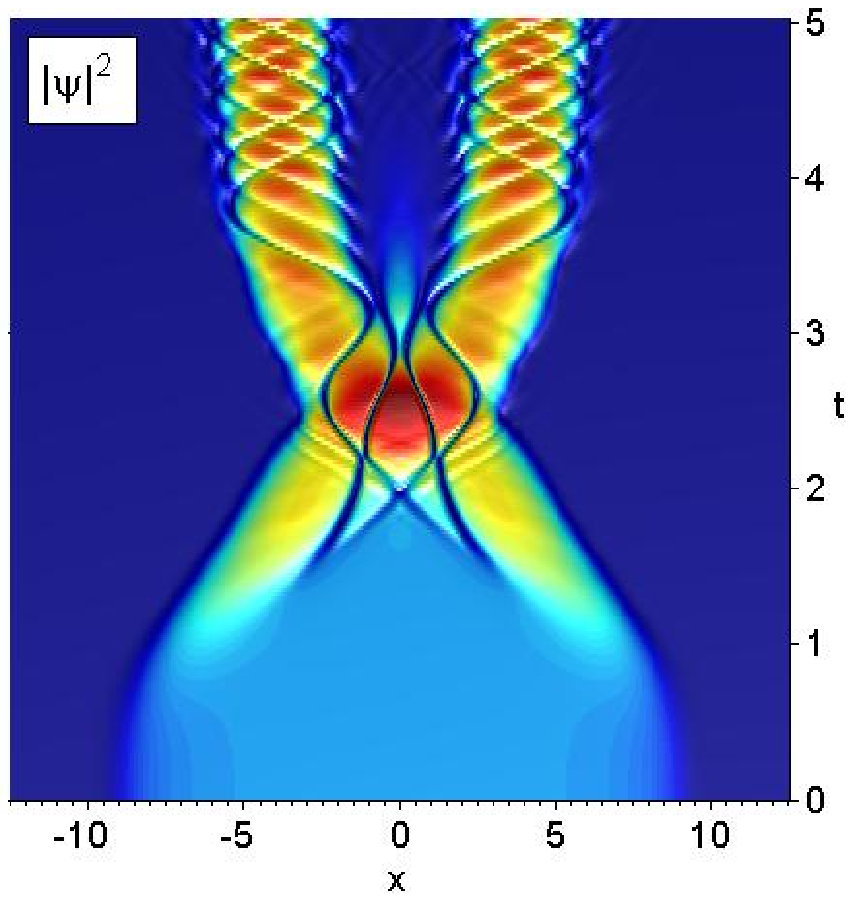}}
\caption{(Color online) Evolution of the wave function according
to Eq. (\protect\ref{quint}) with kernel (\protect\ref{kernel}),
when the strength of the dipolar interaction is ramped up as
$d(t)=2\left[ 1+\mathrm{tanh}(\protect\gamma t)\right] $. Left
panel: When the strength of the dipolar interaction is slowly
raised with $\protect\gamma =0.1$, dark solitons do not emerge,
while the TG gas performs contracting-expanding oscillations,
preserving its integrity. Middle panel: At a moderately fast raise
of $d(t)$, with $\protect\gamma =0.3$, a dark soliton-antisoliton
pair is generated, making the quasi-particle bouncing of the dark
solitons evident. Right panel: Swiftly ramping up $d(t)$ with
$\protect\gamma =0.5$ gives rise to multiple generation of dark
solitons and splitting of the original bright soliton. In time
interval $2<t<3$, which precedes the breakup of the bright
soliton, repeated dark-soliton collisions are observed.
Oscillating dark solitons remain in the split parts of the
original bright one.} \label{fig4}
\end{figure}

The stationary solitons in the model of the dipolar TG gas are
formed due to the balance between the contact repulsion and
long-range attraction between atoms. When the strength of one of
these forces is varied in time, solitons naturally shrink or
expand. The most significant observation following from the
numerical simulations is that, when the strength of the DD
interaction is swiftly ramped up, dark soliton-antisoliton pairs
are created, as shown in Fig. \ref{fig4} (here, ``solitons" and
``antisolitons" are defined as patterns with opposite signs of the
phase gradient across the density depression, see the left panel
in Fig. \ref{fig6}).

The number of the generated dark soliton-antisoliton pairs depends on the
speed at which the dipolar interaction is ramped up. Reaching the edge of
the flat-top background, the dark soliton is reflected back towards the
center. This effect can be seen even after the original soliton splits, see
the right panel in Fig. \ref{fig4}. Colliding at the center, two dark
solitons interact repulsively and bounce back. This is a manifestation of
the particle-like nature of dark matter-wave solitons, which was
experimentally observed in BEC \cite{becker}.

\begin{figure}[tbp]
\centerline{\includegraphics[width=5cm,height=6cm,clip]{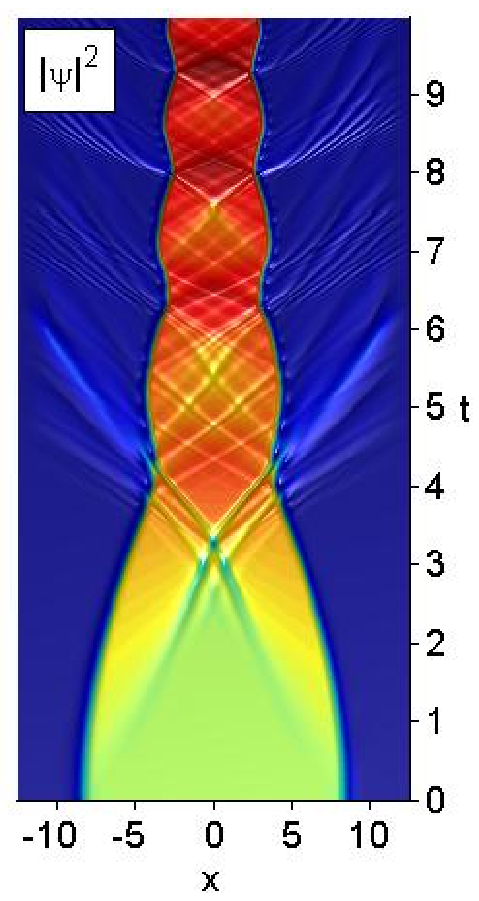}\quad
            \includegraphics[width=5cm,height=6cm,clip]{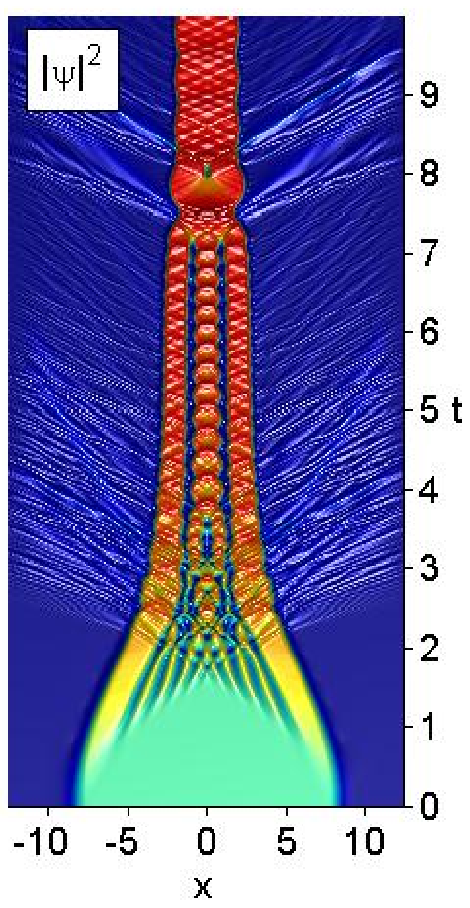}\quad
            \includegraphics[width=5cm,height=6cm,clip]{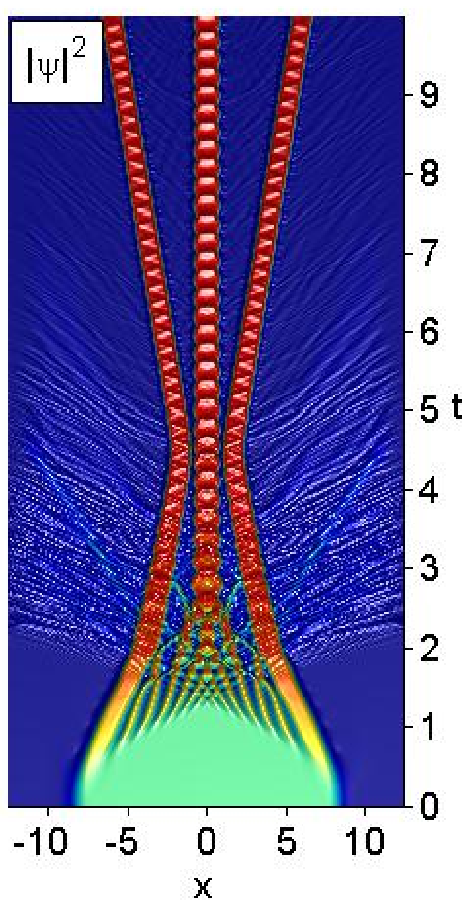}}
\caption{(Color online) Numerical simulations similar to those displayed in
the previous figure, but for local equation (\protect\ref{locgpe}). The
coefficient in front of the cubic nonlinearity is varied in time according
to $\protect\beta (t)=4N[d(t)]^{2}$, where $d(t)$ is the same as in the
previous figure. The initial state is the flat-top soliton of local
cubic-quintic equation (\protect\ref{locgpe}), which is shown in Fig.
\protect\ref{fig3} by the solid red line. In contrast to the situation
observed in the simulations of the nonlocal equation, dark solitons \emph{do
not }appear on top of the flat-top soliton.}
\label{fig5}
\end{figure}

\begin{figure}[tbp]
\centerline{\includegraphics[width=8cm,height=6cm,clip]{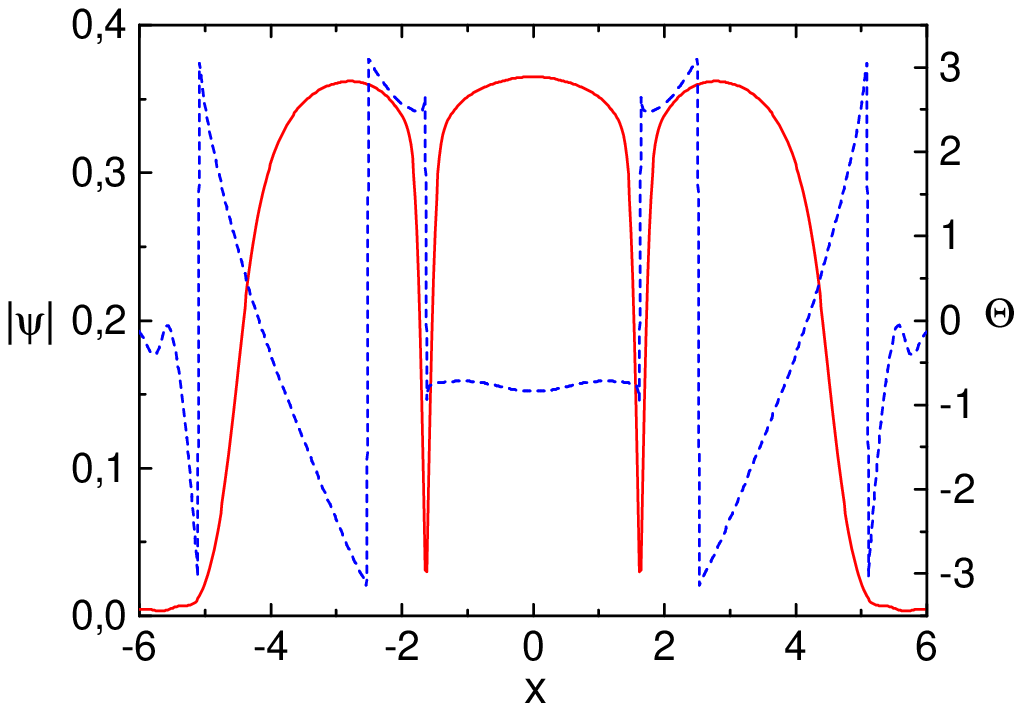}\quad
            \includegraphics[width=8cm,height=6cm,clip]{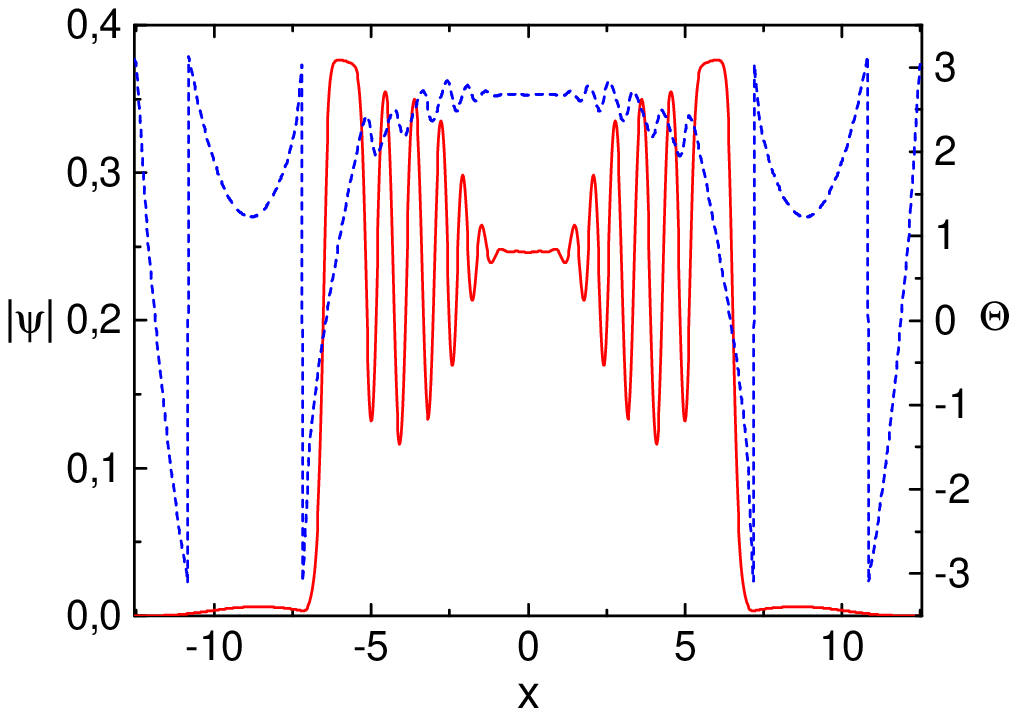}}
\caption{(Color online) Snapshots of the amplitude (red solid
line, left axis) and phase (blue dashed line, right axis) of the
wave function shown in Figs. \protect\ref{fig4} and
\protect\ref{fig5}, at $t=3.5$ (left panel) and $t=1.2$ (right
panel), respectively. Left panel: A phase jump of $\approx
\protect\pi $ across the density dips is the indication that the
emerging excitations are dark solitons. Phase jumps outside the
dips are usual $2\protect\pi $ phase wraps. Right panel: absence
of phase jumps across density modulations implies that these are
quasi-linear excitations, rather than dark solitons.} \label{fig6}
\end{figure}

In order to highlight distinctive features of solitons in the
dipolar TG gas, we have performed numerical simulations, similar
to those shown in Fig. \ref{fig4}, also with the local counterpart
of the original Eq. (\ref{quint}), i.e., Eq. (\ref{locgpe}), as
shown in Fig. \ref{fig5}. In the case of local NLSE, the varying
dipole moment $d(t)$ is emulated by varying the coefficient of the
cubic nonlinearity, $\beta (t)$, in Eq. (\ref{locgpe}). From
comparing Figs. \ref{fig4} and \ref{fig5}, one can observe
drastically different behaviors. The most prominent distinction
concerns the generation of long-lived oscillating dark solitons on
top of the flat-top soliton in the nonlocal model. The other
difference concerns the emission of linear waves under the varying
coefficient of the cubic nonlinearity -- $d(t)$ or $\beta (t)$,
respectively. Namely, the soliton of the local NLSE, Eq.
(\ref{locgpe}), strongly radiates, while the soliton of nonlocal
equation (\ref{quint}) shows almost no radiation. The mode of
splitting of these two types of the solitons, when the coefficient
in front of the cubic nonlinear local term is rapidly varied, is
also very different (see middle and right panels in Figs.
\ref{fig4} and \ref{fig5}).

In experiments performed so far, dark solitons were created by optically
imprinting a phase gradient onto a cigar shaped BEC \cite{carr}. The present
approach, based on ramping up the strength of the DD interactions, may be
considered as an alternative approach to the controllable creation of dark
solitons in quantum gases.

\subsection{Free expansion of the TG gas}

Expansion of a quantum gas released from the potential trap bears important
information about correlations in the system. For instance, the momentum
distribution of bosonic atoms, which expands after sudden removal of the
trapping potential, has been the key evidence showing that the atoms
exhibited a pronounced fermionic behavior, i.e., the TG regime has been
achieved \cite{paredes}. During the expansion of a 1D gas of hard-core
bosons, the momentum-distribution function becomes equal to that of the
equivalent non-interacting fermions. This phenomenon, known as the dynamical
fermionization, is the most interesting manifestation of the Bose-Fermi
duality \cite{rigol}. Quantum correlations in the dynamically evolving TG
gas in an OL were studied in Ref. \cite{pezer}.

Bright solitons in the TG gas with attractive DD interactions can offer
additional possibilities in exploring properties of quantum gases.
Specifically, since in the present setting the strong repulsion between
bosonic atoms is balanced by the long-range DD attraction, the momentum
distribution of a freely expanding gas, achieved by suddenly switching off
or decreasing the strength of the DD interaction, develops from the initial
spatial distribution specific to bright solitons (in contrast to that
corresponding to the harmonic potential or box-shaped trap in previously
studied settings). In the experiment, the DD interaction can be turned off
by the rotation of the external magnetic field, or by eliminating the
external dc electric field responsible for the induced electric dipole
moment.

\begin{figure}[tbp]
\centerline{\includegraphics[width=8cm,height=8cm,clip]{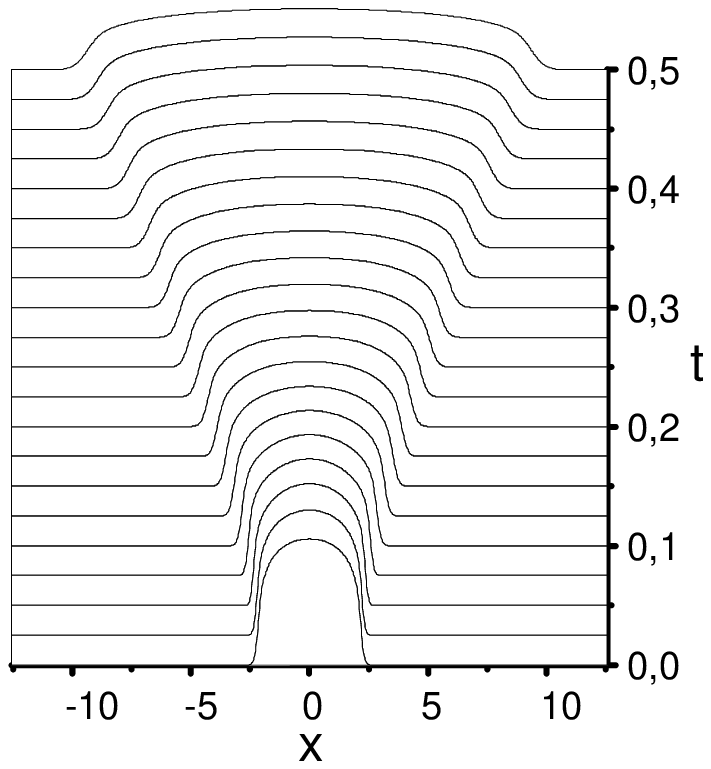}}
\caption{The free expansion of the TG gas after the strength of the dipolar
interaction was suddenly dropped to zero. The initial state is the
compacton-like soliton depicted in Fig. \protect\ref{fig1} by the dashed
line. }
\label{fig7}
\end{figure}

In Fig. \ref{fig7} we illustrate the free expansion of the TG gas from the
compacton-like soliton state shown in Fig. \ref{fig1}. Investigation of the
scaling law governing the expansion of the TG gas of dipolar atoms, as well
as of its momentum distribution, may be interesting topics for subsequent
studies.

\section{Interactions and collisions of dipole solitons}

The wave-particle duality of solitons can be most clearly observed
in their interactions and collisions. While ``genuine" solitons in
integrable models collide strictly elastically, solitons of
non-integrable models show complex collision behaviors, ranging from
almost elastic to fully destructive. Inelasticity of collisions show
up as significant emission of linear waves resulting from the
collision, merger of colliding solitons into a single dynamically
evolving wave packet (including a possible transition to the wave
collapse), splitting of solitons, and multiple generation of
secondary solitons.

Collisions of matter-wave solitons composed of cold atoms with
contact interactions have been extensively studied, both in
numerical simulations \cite{parker} and in real experiments
\cite{martin}. In particular, inelasticity in collisions of
solitons described by the local NLSE\ with self-focusing cubic and
quintic terms, the latter one generated by the deviation of the
effective equation from the one-dimensionality, were analyzed in
Ref. \cite{Lev}.

Collision of matter-wave solitons in BEC with competing cubic
local and dipolar interactions has recently been studied by means
of numerical simulations in Ref. \cite{cuevas}. Here, we explore
collisions between dipole solitons in the TG-gas model. Since the
objective is to reveal features introduced by the long-range
dipolar forces, we focus on the interactions and collisions
between compacton-like solitons, whose properties are dominated by
dipolar forces.

There are essential differences of the present model in comparison
to work \cite{cuevas}: the self-defocusing nonlinear term is here
quintic, and the arrangement of interacting solitons is different.
Namely, in our case one soliton (the ``target") is at rest at the
origin ($x=0$), while the other (the ``missile") is set in motion
towards the target. This setting allows us to explicitly investigate
the momentum exchange between the colliding solitons.

An important conclusion suggested by the numerical experiments is
that the long-range attraction is the dominant force between the
interacting solitons. This nonlocal force is phase-independent and
much stronger than usual short-range interaction forces which
depend on the phase shift between the solitons.

\begin{figure}[tbp]
\centerline{(a)\includegraphics[width=3.5cm,height=7cm,clip]{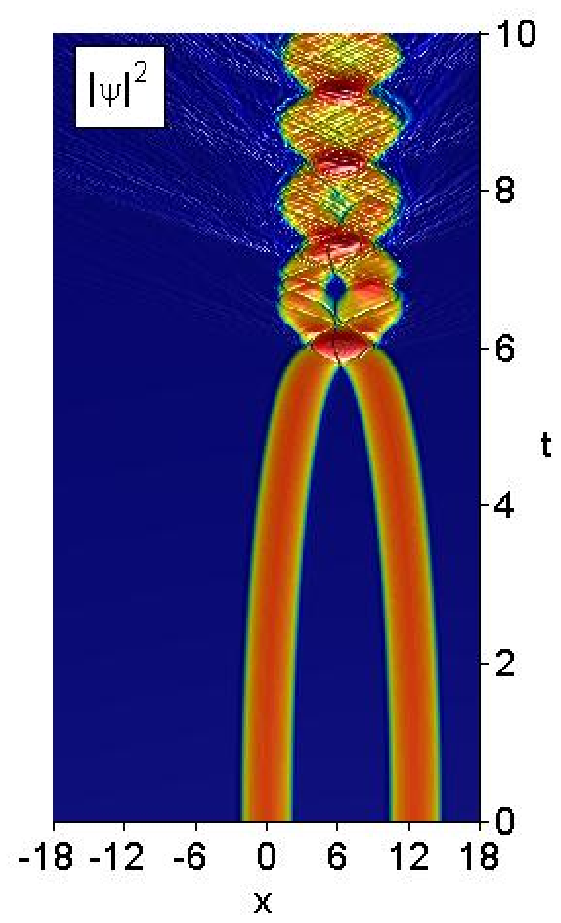}
            (b)\includegraphics[width=3.5cm,height=7cm,clip]{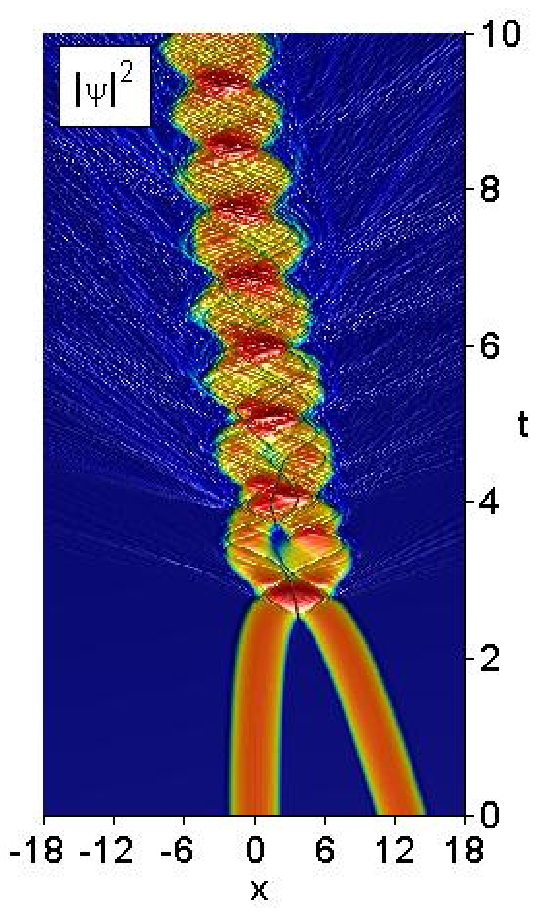}
            (c)\includegraphics[width=3.5cm,height=7cm,clip]{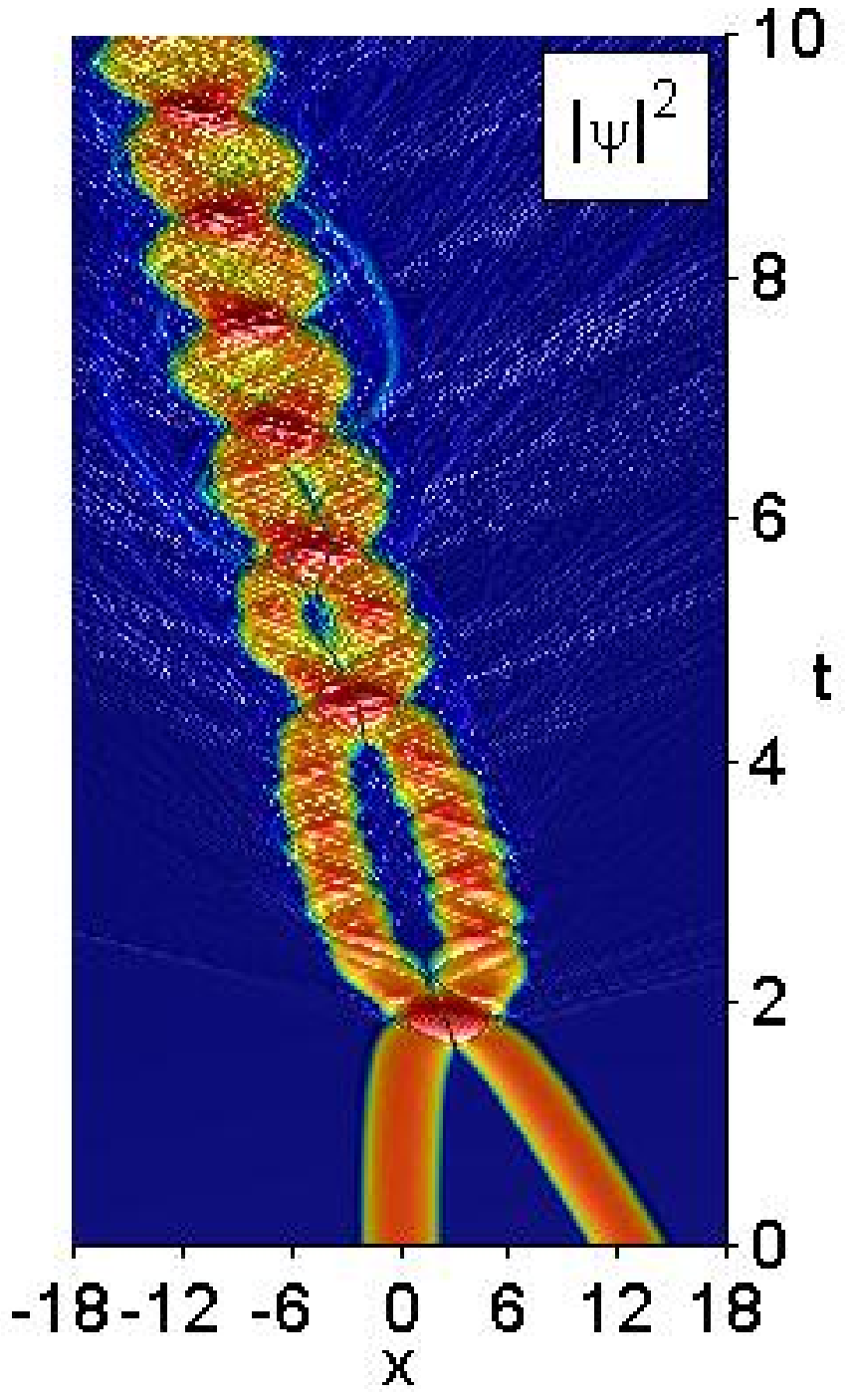}
            (d)\includegraphics[width=3.5cm,height=7cm,clip]{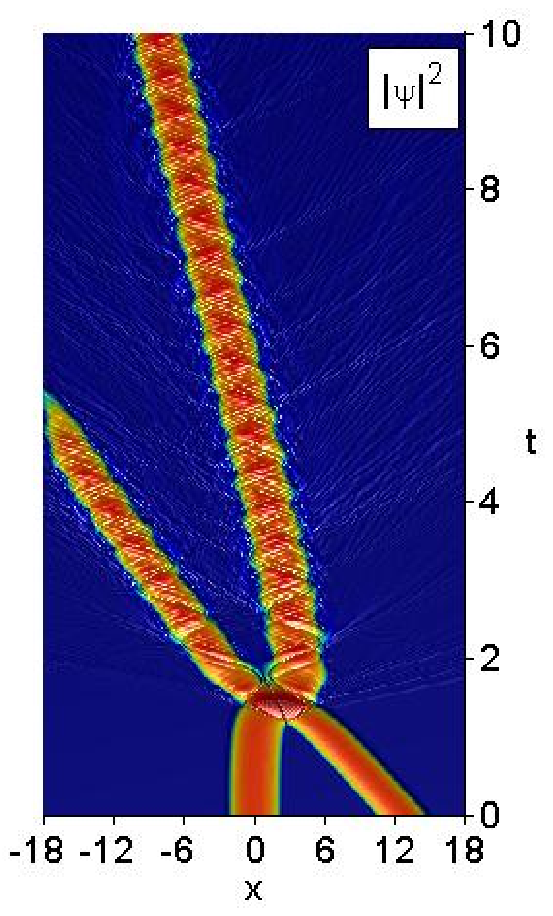}}
\caption{(Color online) Collisions of two compacton-like solitons
(shown in Fig. \ref{fig1} by the dashed line) placed at distance
$\protect\delta x=4\pi$. (a) Two quiescent (zero-velocity) dipole
solitons attract each other and merge into a breather, whose
center-of-mass is still. (b)-(d) When the right soliton is set in
motion with velocity $v_0$ towards the left (quiescent) one, the
outcome of the collision depends on the initial velocity. At
moderate velocities, $v_0=-2$ (b) and $v_0=-4$ (c), solitons merge
into a breather, moving in the same direction as the ``missile"
soliton. At larger velocity $v_0=-6$ (d), the two solitons
separate after the collision. The ``target" acquires the velocity,
while the ``missile" reduces its velocity, in accordance with the
conservation of the total momentum.} \label{fig8}
\end{figure}

In the first set of the numerical experiments, we set two
quiescent (zero-velocity) dipole solitons at distance $\delta
x=4\pi $ from each other, and observed their evolution. In either
case of in-phase (see Fig. \ref{fig8} (a)) and $\pi $ -
out-of-phase (not shown) pairs of the solitons, they attracted
each other and merged into a breather. The fact that the merger
time in both cases was almost the same ($t\simeq 6$) indicates
that the phase-dependent short-range interaction force has played
no tangible role in the dynamics.

Next, we consider collisions between moving solitons. With this
objective in mind, we prepared the initial state with one soliton
set at the origin ($x=0$) with zero velocity ($v=0$), and the
other one placed at $x=4\pi $ with finite velocity $v_{0}$.
General features of soliton collisions can be summarized as
follows (see Fig. \ref{fig8} (b), (c), (d) ): at moderately small
velocities, e.g., $v_{0}=-2$, solitons merge into a single
breather. In fact, the dipolar attraction between the solitons
facilitates the merger. At greater velocities (such as
$v_{0}=-4$), the initially moving soliton passes through the
quiescent one, imparting to it a small velocity in the same
direction. In this case, although the solitons separate after the
collision, the dipolar attraction between them overcomes the trend
to the separation and solitons again merge into a single breather.
At still greater velocities (e.g., $v_{0}=-6$), the two solitons
separate and fly apart overcoming the dipolar attraction.
The conservation of the total momentum of the colliding solitons
can be clearly observed in these numerical experiments.

\section{Conclusions}

In this work, our aim was to study the existence, stability and
basic dynamical properties of bright one-dimensional solitons
based on the balance between the local (hard-core) repulsion and
long-range DD (dipole-dipole) attraction between atoms in the
model of the TG gas. It was found that, depending on the number of
atoms and strength of the DD interaction, the solitons assume
flat-top or compacton-like shapes. For solitons of the former
type, solution of the local cubic-quintic NLSE with competing
nonlinearities is found to be a good approximation. Numerical
simulations of the underlying nonlocal equation with a variable
strength of the DD interaction have revealed various dynamical
regimes, including the formation of dark solitons on top of a
bright one (of the flat-top type), particle-like collisions
between them, splitting of the flat-top solitons, and self-similar
ballistic expansion of the gas after dropping the DD attraction.
Collisions between bright solitons of the compacton type have been
investigated in different regimes. The strong dipolar attraction
between the solitons explains that, in many cases, the colliding
solitons merge into a breather.

\section*{Acknowledgements}

We thank E N Tsoy for valuable discussions and helpful
suggestions. F Kh A and B B B acknowledge partial support from the
Fund for Fundamental Research of the Uzbek Academy of Sciences
under Grant No. 10-08. M S acknowledges partial support from the
Istituto Nazionale di Fisica Nucleare (INFN), Gruppo Collegato di
Salerno, Sezione di Napoli. The work of B A M was supported, in a
part, by the German-Israel Foundation through grant No. 149/2006.

\end{document}